\documentclass{iopart}

\usepackage{iopams}
\usepackage{graphicx}

\begin{document}

\title{Low temperature investigations of single silicon vacancy colour centres in diamond}

\author{Elke Neu$^{1,2}$, Christian Hepp$^1$, Michael Hauschild$^1$, Stefan Gsell$^3$, Martin Fischer$^3$, Hadwig Sternschulte$^{4,5}$, Doris Steinm\"uller-Nethl$^4$, Matthias Schreck$^3$ and Christoph Becher$^1$}

\address{$^1$ Universit\"at des Saarlandes, Fachrichtung 7.2 (Experimentalphysik), Campus E2.6, 66123 Saarbr\"ucken, Germany}
\address{$^2$ Universit\"at Basel, Departement Physik, Klingelbergstrasse 82, 4056 Basel, Switzerland}
\address{$^3$ Universit\"at Augsburg, Lehrstuhl f\"ur Experimentalphysik IV, Universit\"atsstr. 1 (Geb\"aude Nord), 86135 Augsburg, Germany}
\address{$^4$ Komet RhoBeSt GmbH, Exlgasse 20a, 6020 Innsbruck, Austria}
\address{$^5$ Fakult\"at f\"ur Physik - Physik Lehre Weihenstephan und Physik Department E19, Technische Universit\"at M\"unchen, James-Franck-Strasse 1, 85748 Garching, Germany}
\ead{christoph.becher@physik.uni-saarland.de}
\begin{abstract}
We study single silicon vacancy (SiV) centres in chemical vapour deposition (CVD) nanodiamonds on iridium as well as an ensemble of SiV centres in a high quality, low stress CVD diamond film by using temperature dependent luminescence spectroscopy in the temperature range 5--295 K. We investigate in detail the temperature dependent fine structure of the zero-phonon-line (ZPL) of the SiV centres. The ZPL transition is affected by inhomogeneous as well as temperature dependent homogeneous broadening and blue shifts by about 20 cm$^{-1}$ upon cooling from room temperature to 5 K. We employ excitation power dependent  $g^{(2)}$  measurements to explore the temperature dependent internal population dynamics of single SiV centres and infer almost temperature independent dynamics.
\end{abstract}

\submitto{\NJP}
\section{Introduction}
In recent years, single colour centres in diamond have attracted wide research interest due to exciting potential applications such as solid state based single photon sources (for a review see \cite{Aharonovich2011}) or solid state spin qubits \cite{Ladd2010} that can be optically manipulated and read out \cite{Gruber1997,Batalov2008,Santori2006a}. Silicon vacancy (SiV) colour centres in diamond have been identified as a very promising system for the realization of bright, narrow bandwidth, room temperature single photon sources: The fluorescence of single SiV centres is predominantly concentrated in the purely electronic transition, the zero-phonon-line (ZPL), at approximately 740 nm which features a room temperature width of down to 0.7 nm \cite{Neu2011}. Single SiV centres can be feasibly created by ion implantation \cite{Wang2006} or by \textit{in situ} doping  during chemical vapour deposition (CVD) \cite{Neu2011,Neu2011a,Neu2011b}. Moreover, the emission of single SiV centres is fully linearly polarized at room temperature \cite{Neu2011b}, which is advantageous for applications in quantum cryptography \cite{Bennett1984} or in the quantum frequency conversion of single photons \cite{Zaske2011}. However, the spatial structure of the SiV complex as well as its symmetry and the electronic level scheme of the centre are still under debate \cite{Goss1996,Moliver2003}. In previous work \cite{Clark1995,Sternschulte1994}, the SiV centre's ZPL has been observed to split into two doublets at low temperature. This unique four line fine structure pattern may be regarded as a 'spectral fingerprint' of the SiV centre in high quality, low stress diamond and may also be used to unambiguously identify single colour centres as SiV centres.  The origin of the fine structure is not clear at present; in an empirical model it is attributed to a level scheme where excited and ground state are split into two sub-states \cite{Clark1995}. Depending on the assumed spatial symmetry of the SiV complex, the ZPL fine structure has been attributed to a splitting of degenerate electronic eigenstates due to a Jahn-Teller effect or to a tunneling between different spatial realizations of the centre resulting in the splitting of states \cite{Goss1996,Moliver2003}. Moreover, as the spin state of the centre is not clear at present \cite{DHaenens2011}, one might also expect level splittings due to spin-orbit or spin-spin interactions. Due to the unidentified level scheme, the use of SiV centres in quantum information applications requiring detailed knowledge of the level scheme at low temperature as well as polarization of the optical transitions and their linewidth (e.g., \cite{Batalov2008,Santori2006a}) is precluded so far. No detailed studies on low temperature or even temperature dependent luminescence of single SiV centres are reported in the literature. In this work, we use temperature dependent (5--295 K) investigations of single \textit{in situ }created SiV centres as well as SiV centre ensembles to elucidate the ZPL fine structure, line shift and widths. We furthermore investigate the temperature dependent population dynamics of single SiV centres.
\section{Experimental}
\subsection{Samples \label{samples}}
In the present work, we investigate two types of samples. First, to study the fine structure of the ZPL of an SiV ensemble, we employ a single crystal diamond film grown using a modified hot filament CVD technique (sample 1). Homoepitaxial growth on a (001) Ib high pressure high temperature (HPHT) diamond substrate (Sumitomo) eliminates stress arising from thermal expansion mismatch of substrate and diamond and minimizes the contribution of defects like dislocations. Stress is supposed to be the main source of lineshifts and thus inhomogeneous broadening of the SiV centre ZPL \cite{Zaitsev2001}. We furthermore accomplish high crystalline quality, as witnessed by the well defined luminescence properties of the incorporated SiV centres, by applying optimized growth conditions involving a low methane fraction (0.26\% CH$_4$ in H$_2$) and slow growth. To estimate the growth rate and thickness of the homoepitaxial film (sample 1), we fabricate an additional polycrystalline reference sample on a non-diamond substrate using identical growth parameters and nanodiamond seeding. We estimate a growth rate of approximately 10 nm/h equivalent to a total film thickness of 80--100 nm. Note that this polycrystalline reference sample has not been used for any fluorescence investigations. In the homoepitaxial thin film, SiV centres are created \textit{in situ} due to residual silicon contamination of the CVD reactor. The silicon contamination of the reactor might arise due to the use of silicon substrates in previous coating runs. Confocal scans of sample 1 reveal a continuous distribution of SiV fluorescence. Thus, we are not able to optically address single centres in the homoepitaxial film as their area density significantly exceeds 1 centre per $\mu$m$^2$. Therefore, all fluorescence measurements on sample 1 represent the average over the SiV ensemble present in the detection volume of the confocal setup. We point out that the HPHT substrate does not contain SiV centres, thus the observed SiV centre ensemble fully resides in the high quality CVD diamond film. In contrast, nitrogen vacancy centres natively occur in the underlying HPHT substrate. Due to the low thickness of the homoepitaxial CVD film, we cannot distinguish between nitrogen vacancy centres in the HPHT substrate and in the CVD film. To achieve single emitter addressing in homoepitaxial thin films, the use of a cleaner reactor system is recommended. Furthermore, thinner films or the fabrication of nanostructures, e.g., nanowires \cite{Babinec2010} may promote the detection of single centres.

\begin{figure}[h]
\centering
\includegraphics[width=0.75\textwidth]{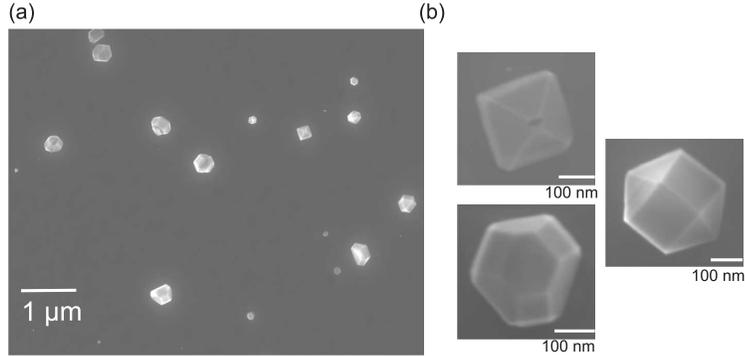}
\caption{SEM images of CVD grown NDs on sample 2b: (a) overview image displaying spatially isolated NDs (b)  detailed images of well faceted NDs.  \label{fig_SEMsample2}}
\end{figure}

Second, we employ spatially isolated nanodiamonds (NDs) grown using microwave plasma CVD on an iridium (Ir) surface to investigate single SiV centres. The 150 nm thick Ir layer is deposited onto a thin buffer layer of yttria-stabilized zirconia on a silicon wafer \cite{Gsell2004}. Randomly oriented NDs are grown on seed diamonds (Microdiamant, MSY GAF 0--0.03 $\mu$m) which are distributed on the substrate by spin coating from an aqueous solution (for details of the seeding process see \cite{Neu2011}). We use two ND samples: Sample 2a contains NDs with an average size of 130 nm (growth conditions: 25 min, 1\% CH$_4$ in H$_2$, 2000 W microwave power, 30 mbar); Sample 2b contains NDs with an average size of 220 nm (growth conditions: 55 min, 0.4\% CH$_4$ in H$_2$, 2000 W microwave power, 30 mbar). Scanning electron microscope (SEM) images of sample 2a can be found in \cite{Neu2011}; figure \ref{fig_SEMsample2} displays SEM images taken from sample 2b.  By adjusting the seed density, we obtain approximately 2 NDs per $\mu$m$^2$ on sample 2a and approximately 0.3 NDs per $\mu$m$^2$ on sample 2b. In both samples, single SiV centres are created \textit{in situ} due to plasma etching of the underlying silicon substrate at its uncovered edges and subsequent silicon incorporation. As discernible from figure \ref{fig_SEMsample2} and reference \cite{Neu2011}, both samples display highly facetted NDs featuring shapes close to the cubo-octahedral shapes of single crystal diamonds. The growth conditions employed for sample 2b seem to be superior compared to the conditions used for sample 2a as in the SEM images the NDs are found to exhibit even cleaner surfaces without cracks and distortions. In contrast to the homoepitaxial film, individual NDs, in general, exhibit a variable stress state that develops during the growth (for details see, e.g., \cite{Neu2011,Neu2011b}). The NDs on Ir have been grown using methane with a purity of 99.9995\% and hydrogen of initially 99.999\% which was further purified by a silver palladium membrane in order to reduce the nitrogen content by several orders of magnitude. As a result, no nitrogen vacancy centre luminescence could be detected. The nitrogen electron-paramagnetic-resonance signal in bulk crystals formerly grown under these conditions was below the detection threshold ($\approx$40 ppb).
\subsection{Experimental Setup \label{setup}}
To analyze the colour centre fluorescence, we employ a homebuilt confocal laser microscope. For the temperature dependent single emitter investigations, a flow cryostat (Janis Research, ST-500LN) operated with liquid helium is integrated into the optical setup. Temperatures between approximately 5 K and room temperature (295 K) can be selected via resistive electrical heating of the cryostat sample mount. The temperatures given throughout this paper have been measured at the sample mount of the cryostat. The excitation laser, either a frequency doubled diode pumped solid state laser (at 671 nm or 532 nm) or a tunable Ti:sapphire laser (here 690 nm), is focused through a microscope objective with NA=0.8 (Olympus, magnification 100x). Using a dichroic beam splitter and dielectric filters, we separate the SiV fluorescence in the spectral range 730--750 nm from residual laser light reflected from the sample surface. The fluorescence is coupled into a multimode fibre. Subsequently, the fluorescence is directed to a Hanbury Brown Twiss (HBT) setup (avalanche photodiodes, APDs: Perkin Elmer SPCM-AQRH-14, correlation electronics: Pico Quant, Pico Harp 300) to measure the intensity autocorrelation (g$^{(2)}$) function of the fluorescence. Alternatively, the fluorescence can be analyzed spectrally. We first use a grating spectrometer (Horiba Jobin Yvon, iHR 550). For cryogenic temperature spectroscopy, we employ a grating with 1800 grooves/mm enabling an approximate resolution of 0.05 nm (30 GHz) at 738 nm.

To perform emission spectroscopy with higher spectral resolution the setup is also equipped with a scanning Fabry-Perot interferometer (FPI). The FPI exhibits a free spectral range (FSR) of 34 GHz in a planar mirror configuration and a finesse $F$ of 54 (yielding a resolution of 0.6 GHz). The comparably low finesse was chosen intentionally in favour of a high signal transmission of $T = 86 \%$. We preselect a single line out of the four line fine structure  using a silica etalon with dielectric coating (reflectivity $R =94 \%$) which acts as a 44 GHz bandpass filter. The preselection is necessary as the fine structure of the ZPL spans a spectral range much larger than the FSR of the FPI. The transmitted fluorescence is detected using the APDs of the HBT setup.
\section{Results}
\subsection{Photoluminescence spectra \label{PLspectra}}
\begin{figure}[h]
\centering
\includegraphics[width=0.8\textwidth]{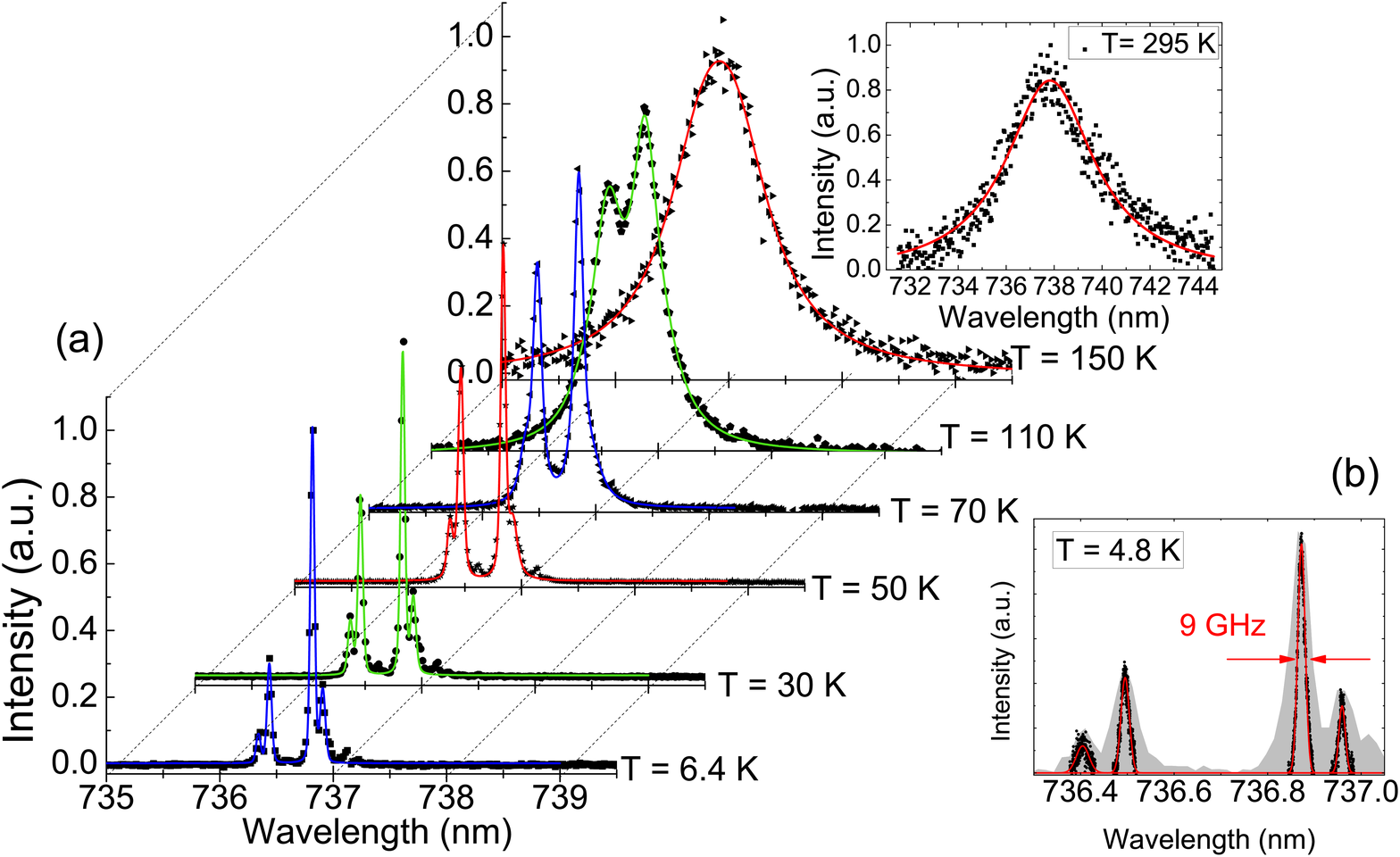}
\caption{Fluorescence of SiV centre ensemble in a homoepitaxial CVD diamond film (sample 1) (a) Temperature dependence of the ZPL fluorescence excited with 532 nm laser light. Spectra up to 150 K allow for the observation of the temperature dependent variations of the fine structure. The inset shows the ZPL at room temperature for comparison, note the different wavelength scales. (b) High resolution spectrum of the fine structure lines at 4.8 K recorded using a scanning FPI (black dots, excitation 690 nm). Note that the spectrum shown here is composed of the measurements of the individual fine structure lines preselected as discussed in section \ref{setup}. The curves have been fitted using Gaussian peak functions (red solid lines) to yield the optimal correspondence with the measured data. The underlying grey spectrum, acquired with the grating spectrometer, serves as a guide to the eye to illustrate the correspondence between the FPI and spectrometer measurements.    \label{Sivtempscd}}
\end{figure}
First, we investigate the low temperature photoluminescence spectra of ensembles of SiV centres in sample 1 which contains a high density of SiV centres. As a preliminary remark, we note that in polycrystalline diamond films on silicon (sample description see, e.g., \cite{Neu2011a}), we recently found a temperature independent, inhomogeneous ZPL width of approximately 2 nm that easily dominates a ZPL fine structure. This observation is in accordance with the literature \cite{Feng1993,Clark1995} and illustrates the sensitivity of the SiV ZPL wavelength to the local environment. In contrast, homoepitaxial CVD diamond films as introduced in section \ref{samples} (sample 1) are a host material with high crystal quality and low stress level enabling the observation of the fine structure of the ZPL of an SiV ensemble as displayed in figure \ref{Sivtempscd}(a): At temperatures below 70 K, the four line fine structure develops, accompanied by additional peaks stemming from less abundant silicon isotopes (Si$^{29}$, Si$^{30}$) in accordance with the literature \cite{Clark1995,Sternschulte1994}. This fine structure has been explained using an empirical level scheme with a split ground and excited state \cite{Clark1995}. We determine the (larger) excited state splitting to be 260 GHz, whereas the ground state splitting amounts to 50 GHz, thus coinciding with observations in the literature within 10\%. Upon cooling, the ZPL blue shifts by approximately 1 nm (18.4 cm$^{-1}$, 2.3 meV) reproducing the observation in \cite{Feng1993} within 20\%. The observed linewidth of the luminescence spectrum at low temperature is limited by the resolution of the employed grating spectrometer.
To further investigate the linewidth, we thus employ high resolution emission spectroscopy using a scanning FPI. The obtained data are shown in figure \ref{Sivtempscd}(b) for each of the four fine structure lines. The measured linewidths range between 9 GHz and 15 GHz. Similar values have been reported for the inhomogeneously broadened ZPL of an ensemble of NV centres in high-quality CVD-grown diamond \cite{Santori2010}. As the inhomogeneous distribution of ZPL wavelengths is, in general, larger for the SiV centre than for the NV centre (see e.g. \cite{Neu2011b}) the linewidth measurements obtained here indicate the high structural quality of the homoepitaxial diamond layer.

\begin{figure}
\centering
\includegraphics[width=\textwidth]{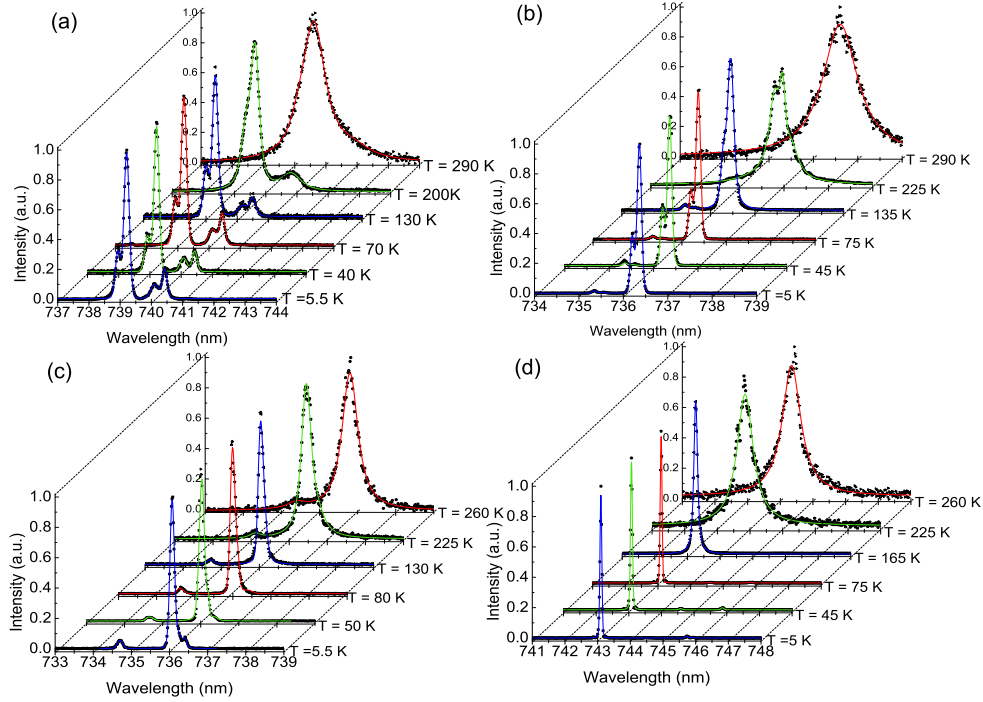}
\caption{Temperature dependent photoluminescence spectra of single SiV centres under 671 nm laser excitation: (a) emitter C1 (0.32 mW excitation power), (b) C2 (0.16 mW), (c) C3 (0.06 mW below 30 K, 0.6 mW above 30 K) and (d) C4 (0.1 mW below 10 K, 0.07 mW above 10 K). \label{PLspektren}}
\end{figure}
In spatially isolated CVD nanodiamonds on Ir, investigation of single, bright SiV centres is feasible \cite{Neu2011,Neu2012a}. We here discuss four emitters in detail, labeled C1--C4. Emitter C1 is hosted by a ND on sample 2a, the other emitters are found in sample 2b. Figure \ref{PLspektren} summarizes temperature dependent luminescence spectra of the investigated emitters. Emitter C1--C3 show a four line fine structure, partially with badly resolved components, while emitter C4 displays a spectrum dominated by a single line. Prior to discussing the spectra in detail, we address the question wether these multi-line spectra, considering the significantly varying brightness of single SiV centres \cite{Neu2012}, can be clearly identified as single emitter spectra. Section \ref{population_dyn} discusses $g^{(2)}$ measurements of the SiV fluorescence under continuous laser excitation. We here anticipate some of the results: the $g^{(2)}$ measurements indicate the presence of single emitters; the measured deviations from $g^{(2)}(0)=0$ are almost fully explained by the timing jitter of the measurement setup and in some cases luminescence background. However, we find residual deviations $<0.15$. These further deviations from the ideal case possibly originate from the presence of a second weak emitter. Assuming that a residual of $g^{(2)}(0)$ of 0.15 is unambiguously detectable, we calculate that it is feasible to identify emitters with an order of magnitude difference in their brightness. Examining the relative line intensities, $g^{(2)}$ measurements for emitter C1 and C3 cannot exclude that the weakest line component originates from a second emitter. For emitter C2, the same holds for the shorter wavelength doublet. However, the following analysis of the fine structure pattern also aids in assigning the observed lines to single SiV centres and their level schemes.

The observed single emitter spectra display three major differences compared to the spectrum of the SiV ensemble in the homoepitaxial CVD diamond film:
First, the absolute positions as well as the spacings and the relative intensity of the fine structure components are altered and vary between different emitters. Second, the room temperature linewidths are significantly smaller for the single emitters. Third, the lines at low temperature display additional broadening. The change of the room temperature linewidth will be addressed in section \ref{homogeneous_width}; the additional line broadening will be discussed in section \ref{sec_spectrdiff}. The varying spectra arise from changes in the local environment of the individual colour centres: Stress in the diamond lattice generally changes the energy levels of colour centres. Symmetry conserving stress changes the absolute energetic position of the levels without affecting the level splittings, whereas symmetry breaking stress changes the level splittings \cite{Grazioso2011}. For SiV centres, the splittings $\Delta E_{e}$, $\Delta E_{g}$ of the excited and ground state, respectively, have been shown to vary under externally applied stress in $\langle100\rangle$ direction: for a stress of 0.4 GPa, $\Delta E_{e}$ amounts to 4.5 times $\Delta E_{e}$ without stress, whereas $\Delta E_{g}$ even increases to a value of 11 times $\Delta E_{g}$ without stress \cite{Sternschulte1994}.  These changes lead to altered positions and spacings of fine structure components and are accompanied by relative intensity changes of the fine structure components \cite{Sternschulte1994}. Particularly, already under a moderate stress of 0.33 GPa in $\langle100\rangle$ direction, a single line dominates the spectrum while other fine structure components mostly vanish \cite{Sternschulte1994}. Thus, even the strongly modified spectrum of emitter C4 can be attributed to an SiV centre with a ZPL fine structure altered by local stress fields. We note that we also observe emitters where only a doublet line is present in the spectrum. For emitter C2 (C1), the fine structure pattern is comparable to the pattern of the ensemble of SiV centres in homoepitaxial diamond: Despite the fact that $\Delta E_{g}$ is twice (three times) the spacing for the ensemble, the ratio ${\Delta E_{e}}/{\Delta E_{g}}$ matches the ensemble case within 6\% (22\%). The results in \cite{Sternschulte1994} document a change of ${\Delta E_{e}}/{\Delta E_{g}}$ for  $\langle100\rangle$ stress. However, no information on the response of SiV centres to stress in other directions is found in the literature. Furthermore, the stress distribution inside the investigated NDs is unknown. We thus attribute the variety of observed line patterns, for which the discussed emitters are just examples, to the local stress states experienced by the individual colour centres. Moreover, static electric fields due to charges in the crystal might influence the spectrum. Data on the Stark shift of the SiV centre is not available in the literature.

\begin{figure}[h]
\centering
\includegraphics[width=0.7\textwidth]{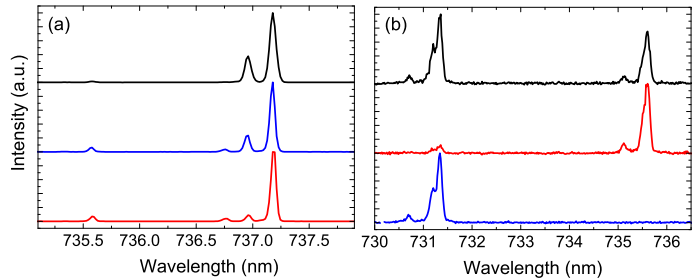}
\caption{Low temperature spectra for different emitters demonstrating instabilities [(a) 8 K, (b) 30 K].
 All spectra are normalized to the peak value and displaced for clarity.  (a) spectrum of an emitter with resolution limited lines, but unstable relative intensity (integration time 60 s blue and red curves, 180 s black curve).  (b) emitter with spectral jumps, but similar spectral pattern in both states (integration time 5 s).   \label{fig_varspec}}
\end{figure}
Whereas many of the SiV centres show stable emission, some emitters with unstable emission spectrum were observed. Figure \ref{fig_varspec}(a) displays an emitter showing narrow, spectrometer resolution limited lines. Remarkably, however, the relative intensity of the fine structure lines varies over time. Figure \ref{fig_varspec}(b), in contrast, displays spectra taken from an emitter undergoing spectral jumps. Each spectrum has been integrated for 5 s and witnesses spectral jumps on a timescale of seconds: Spectra revealing one (both) of the ZPLs arise if the emitter resides in one (both) of the states during the spectrum integration time. Remarkably, the spectra in both states resemble a very similar ZPL fine structure despite the significant wavelength difference. The jumps might occur due to the transfer of an electron between SiV centres under (significantly) different stress conditions if luminescence is only observed from the momentarily charged centre. Spectral jumps have also been reported in \cite{Siyushev2009} and \cite{Muller2011} for colour centres emitting around 770 nm and chromium-related centres. In accordance with \cite{Neu2012a,Shen2008,Siyushev2009}, we furthermore observe blinking or bleaching of single centres. The emission instabilities as well as the additional broadening, commonly referred to as spectral diffusion, might be connected to charged defects or fluctuating charges inside or at the surfaces of the NDs:
\subsubsection*{Ionization of other impurities inside the NDs}\label{spectrdisc}
A prevalent impurity in diamond is substitutional nitrogen $[N_s]$ which forms a deep electron donor with an ionization energy of approximately\ 1.7 eV (729.3 nm, e.g., \cite{Tamarat2006}). Thus, nitrogen donors in the focal volume of our 690 nm, 671 nm or 532 nm excitation laser might be photoionized, forming $[N_s]^+$ and electrons in the conduction band. Additionally, very recent experiments show that $[N_s]$ can also act as an acceptor \cite{Ulbricht2011}.  Due to these processes, fluctuating charges in the conduction band of diamond are created and shift colour centre transitions via the Stark effect \cite{Tamarat2006,Jelezko2002}. For single NV centres, linewidths of up to 90 GHz in nitrogen rich type Ib diamond have been observed \cite{Tamarat2006}. As the $[N_s]$ content of the NDs has been estimated to be below $\approx$ 40 ppb (see section \ref{samples}), we assume that only a few nitrogen atoms -- if any -- are incorporated in NDs with a size of 130 nm.
For the homoepitaxial film, we cannot estimate the nitrogen content. However, it is supposed to be significantly higher than for the CVD NDs. It should be noted that the Ib HPHT diamond substrate used for the growth of the homoepitaxial film as well as the HPHT seed diamonds used for the ND growth contain a significantly higher concentration of $[N_s]$ (in the order of 100 ppm, \cite{Zaitsev2001}).
As a further effect, other impurities in diamond exist in multiple charge states and switch their charge state continuously under laser excitation (photochromism). Charge state conversion has been observed for NV centres \cite{Gaebel2006b} as well as for SiV centres \cite{DHaenens2011}. Thus, also the charging or discharging of other colour centres may contribute to a fluctuating charge background and might lead to line broadening or unstable emission. It remains unclear, why the ensemble in the presumably less pure homoepitaxial diamond is less prone to line broadening.
\subsubsection*{Trapping and release of charges on the surfaces of NDs}
Recent work on 5 nm detonation NDs identifies surface charge traps due to impurities, dislocations and dangling bonds as the major cause of fluorescence blinking of NV centres \cite{Bradac2010}. Similarly, NV centres implanted close to the surface of CVD diamond show strong spectral diffusion compared to \textit{in situ} formed centres deeper in the same sample \cite{Fu2010}. These findings suggest that also light induced fluctuating charges \cite{Wolters2013} at surfaces induce spectral diffusion and non-stable emission. However, we point out that the SiV centres in the 80--100 nm thick homoepitaxial film also experience a close proximity to the surface without suffering from spectral diffusion.
\subsubsection*{Influence of the Ir substrate}
In addition to free charges created in the diamond, one might think of charge transfer between the diamond and the metal
substrate. Ir(001) has a workfunction of 5.67 eV \cite{HCP1995}. Assuming a hydrogen terminated diamond surface (band diagram see \cite{Hauf2011}), electrons from the valence band in diamond might be transferred to unoccupied states of the Ir metal by visible light (density of states for Ir see \cite{Noffke1982}) and thermalize back to the valence band of diamond. Thus, also the light induced charge transfer between substrate and diamond might add to a fluctuating charge background und thus to spectral diffusion of SiV centres.
\subsection{Temperature dependent emission linewidth \label{sec_linewidht}}
The emission linewidth of single colour centres is a crucial parameter considering their application in quantum information processing, in particular considering the generation of indistinguishable photons. The spectra of figure \ref{PLspektren} reveal a temperature dependent broadening of the ZPL. In general, both inhomogeneous and homogeneous line broadening occur, where the latter relates to dephasing processes due to phonons \cite{Siyushev2009}. The temperature dependent homogeneous broadening is a result of quadratic electron phonon coupling and leads to a Lorentzian lineshape $I_L(\lambda)$ \cite{Maradudin1966,Davies1981}.
Especially for single emitters in NDs, spectral diffusion can induce additional broadening. This mechanism constitutes an inhomogeneous broadening and results in a Gaussian emission line shape $I_G(\lambda)$ \cite{Siyushev2009}.
If the charge fluctuations discussed above are temperature independent, one would assume spectral diffusion to be temperature independent, as observed for NV centres in \cite{Wolters2013} in the temperature range 5--20 K and in \cite{Muller2012} for chromium related centres.

If both line broadening types occur the emission line shape is given by the convolution of a Gaussian and a Lorentzian line. This line shape is termed a Voigt-profile $I_V(\lambda)$ \cite{Kirillov1994}
\begin{equation}
I_V(\lambda)=\frac{2A\ln(2)}{\pi^{\frac{3}{2}}}\frac{w_L}{w_G^2}  \int_{-\infty}^{\infty} \frac{e^{-t^2}}{\biggl( \frac{\sqrt{\ln(2)}w_L}{w_G}\biggr)^2+\biggl(\frac{\sqrt{4\ln(2)}(\lambda-\lambda_c)}{w_G}-t\biggr)^2}\, dt
\end{equation}
$w_G$ and $w_L$ are the Gaussian as well as Lorentzian full widths at half maximum (FWHM) of the emission line, $\lambda_c$ is the peak wavelength and $A$ is proportional to the area of the line. By fitting a Voigt-profile to the measured line shapes, we separate homogeneous and inhomogeneous line broadening contributions for the ZPL. The resulting curves are displayed in figure \ref{PLspektren} and match the measured data very well. Note that for the following discussion we have corrected the obtained linewidths for the spectrometer response function (given also by a Voigt profile). A general trend considering the line broadening is observed: For the SiV centre ensemble in homoepitaxial CVD diamond (sample 1), at 5 K linewidths of below 20 GHz have been found (see section \ref{PLspectra}). For emitters C1--C4 in contrast, figure \ref{PLspektren} reveals significantly broadened lines (20--150 GHz) at around 5 K with lineforms close to a Gaussian shape. For elevated temperatures, the observed lineform changes and is closer to a Lorentzian line.

Considering the evaluation of the data, the complex spectral pattern together with the large number of free parameters for each Voigt-profile renders least--chi$^2$ fitting of the emission spectra extremely challenging. Therefore, we have adapted the following strategy to fit the line patterns: At low temperature, we assume $w_L$ to be far below the spectrometer resolution; we thus fix $w_L$ to the value obtained from the spectrometer lineform. The fits result in a constant value for $w_G$ up to about 50 K. After correction for the Gaussian contribution of the spectrometer response, $w_G$ gives the spectral diffusion linewidth for the respective line. We assume $w_G$ to be temperature independent as shown, e.g., in \cite{Muller2012} for chromium related centres. We thus fix $w_G$ to the (mean) value obtained for each line at low temperature and fit $w_L$ representing the increasing homogeneous broadening for higher temperature.
\subsubsection{Homogeneous linewidth \label{homogeneous_width}}
The temperature dependent homogeneous linewidth is determined by the phonon spectrum of the perturbed host crystal, i.e., the crystal including the colour centre, as well as the population of the phonon modes (for details see \cite{Maradudin1966}). The phonon spectrum of the perturbed crystal cannot be directly measured, however, the line broadening can be approximately related to the one-phonon assisted absorption spectrum of the colour centre \cite{Maradudin1966}. Measuring the latter is not feasible in our experiment. Furthermore, deriving the absorption spectrum from the emission spectrum is precluded due to a lack of inversion symmetry of the SiV spectrum about the ZPL \cite{Iakoubovskii2001}. We thus cannot use this approach for a qualitative analysis of the line broadening.

Treating the crystal in Debye approximation, Maradudin derives a $T^7$ dependence of the linewidth for low temperatures if only non-degenerate electronic states take part in the transition \cite{Maradudin1966}. For degenerate electronic states, phonon interactions can induce transitions between degenerate sublevels of  electronic states (dynamic Jahn-Teller effect). Thus, in the case of degenerate electronic states, phonon interactions can result in additional dephasing mechanisms. Hizhnyakov et al.\ extend the theory to degenerate electronic states and include the effect of a softening of bonds (rearrangement of electrons) and thus an enlarged low energy local phonon density in the excited state of the colour centre \cite{Hizhnyakov2002}. They find a $aT^3+cT^7$ ($a,c=const.$) dependence and verify for an NV centre ensemble a dominant influence of the $T^3$ dependence. On the other hand, Hizhnyakov and Reineker \cite{Hizhnyakov1999} attribute a $T^3$ dependence to fluctuating fields which are created as a phonon modulates the distance between the colour centre and other defects in the crystal. The authors emphasize that the $T^3$ dependence has been observed in systems exhibiting a significant inhomogeneous broadening of ensembles, thus indicating an influence of the purity of the host material on the homogeneous broadening. Using perturbation theory, a $T^5$ dependence of the linewidth at low temperature in the presence of a dynamic Jahn-Teller effect has been derived \cite{Hizhnyakov2002}. In principle, taking into account the discussion above, it should be possible to identify the main homogeneous line broadening mechanism from the dominant term in the polynomial $aT^3+bT^5+cT^7$ ($a,b,c=const.$) dependence of the linewidth on temperature. Nevertheless, observations for NV centres in the literature illustrate that the results are not at all unambiguous: Either dominant $T^3$, $T^5$ or $T^7$ broadening has been reported \cite{Hizhnyakov2002,Lenef1996a,Davies1974,Fu2009a}.
\begin{figure}[h]
\centering
\includegraphics[width=\textwidth]{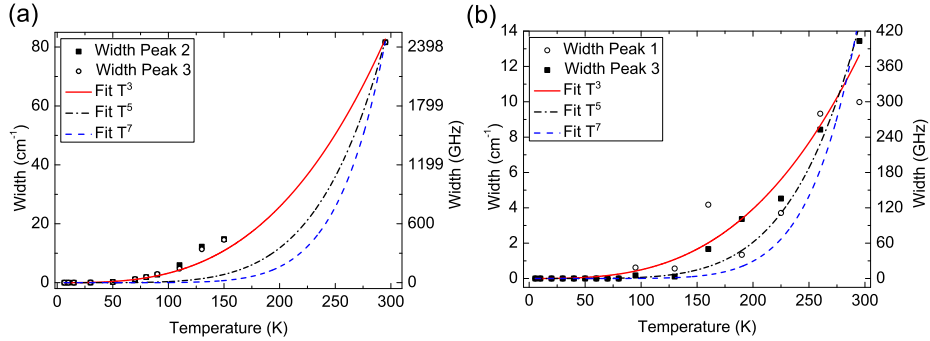}
\caption{Homogeneous linewidth of (a) the SiV ensemble in a homoepitaxial CVD film (sample 1) as well as (b) selected single emitter (C3). The main peaks are used for the analysis as they can be observed over a wide range of temperature. Note that the data points between 150 K and 300 K in (a) are missing due to technical problems during the measurements. The peaks are consecutively numbered with increasing wavelength. For the ensemble (single emitter) the fits take into the account the data from peak 2 (peak 3). \label{homlinew}}
\end{figure}
Figure \ref{homlinew}(a) displays the temperature dependent homogeneous linewidth for the ensemble of SiV centres in the homoepitaxial CVD diamond film; Figure \ref{homlinew}(b) shows the same for emitter C3 as an example for a single emitter. A difference in room temperature linewidth is clearly visible. Fitting the temperature dependent linewidth to either a $T^3$, $T^5$ or $T^7$ dependence, it is clear from figure \ref{homlinew} that the latter does not at all describe the experimental data. For the ensemble in the single crystal as well as emitter C1, C3 and C4, the $T^3$ dependence gives the lowest chi$^2$. While only for emitter C2, $T^5$ leads to the best fit. Taking into account the discussion above as well as the observation of a significant inhomogeneous line broadening (spectral diffusion) we tentatively attribute the main line broadening mechanism as related to defect induced broadening. We note that this observation is similar to very recent results obtained using single chromium related colour centres \cite{Muller2012}.

\subsubsection{Spectral diffusion \label{sec_spectrdiff}}
\begin{figure}[h]
\centering
\includegraphics[width=0.5\textwidth]{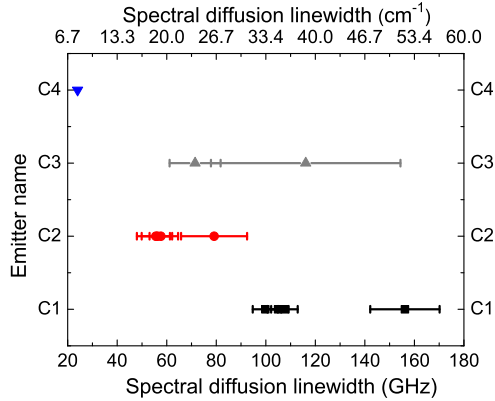}
\caption{Spectral diffusion linewidths of single SiV centres in NDs. The errors reflect the standard deviation of the width obtained at low temperature. As discussed in section \ref{sec_linewidht}, we use low temperature values to estimate the spectral diffusion linewidth up to the temperature where a systematic increase in homogeneous linewidth is found (C1: 110 K, C2: 45 K, C3: 80 K, C4: 10 K).   \label{spzusammenf}}
\end{figure}

For the majority of emitters in NDs, we observe additional inhomogeneous broadening attributed to spectral diffusion at low temperatures.
Figure \ref{spzusammenf} summarizes the Gaussian spectral diffusion linewidths of emitters C1-C4. For emitter C3, only the short wavelength peak and the main peak of the red shifted doublet are taken into account. Note that badly resolved peaks and an unclear peak structure lead to significant errors for this emitter. For emitter C4, we consider only the main peak. Emitter C4 noticeably shows the lowest spectral diffusion linewidth, despite the strongly modified spectral pattern. For all emitters, the measured linewidths vary between 25 GHz and 160 GHz. Significant differences between individual lines of the same emitter may occur with linewidth differences up to nearly 50\%, indicating that different electronic transitions can involve different susceptibility to spectral diffusion. The difference between individual emitters indicates a strong dependence on the local environment, i.e., proximity to possibly ionized impurities or surfaces as also reported for NV centres in NDs \cite{Shen2008}.
From the measurements discussed in section \ref{population_dyn}, we estimate the excited state lifetime of emitters C1, C3, C4 to range from 0.72--1.28 ns yielding a lifetime limited linewidth $\Delta \nu_{FT}$ of 0.12--0.22 GHz. Thus, for emitter C1 and C3, the smallest spectral diffusion linewidth corresponds to $450\cdot \Delta \nu_{FT}$ and $170\cdot \Delta \nu_{FT}$ for emitter C4. We note that we also observe spectrometer resolution limited lines of single SiV centres in CVD NDs [see figure \ref{fig_varspec}(a)]. Furthermore, we find a linewidth of approximately 10 GHz for an SiV ensemble in a high quality, homoepitaxial CVD film (sample 1, see section \ref{PLspectra}). This indicates that narrow lines can be obtained for SiV centres in low-defect environments. We suggest that the main cause of spectral diffusion is the non-resonant excitation laser as indicated by the following observation and also reported for NV centres in the literature, e.g., in \cite{Wolters2013,Robledo2010,Bernien2012}. For emitter C4, the inhomogeneous linewidth changes from 20 GHz at 0.3 times the saturation power $P_{Sat}$ to 55 GHz  at 14 $P_{Sat}$. We thus suggest that by using resonant excitation, reducing the laser photon energy as well as the necessary excitation power, lower linewidths are feasible for single SiV centres in NDs. As all single emitters have been investigated under 671 nm laser excitation only and resonance fluorescence detection was unavailable with the employed setup, the investigation of laser wavelength dependent linewidth broadening remains subject to future studies.
\subsection{Temperature dependent ZPL shifts \label{ZPL_shift}}
\begin{figure}[h]
\centering
\includegraphics[width=\textwidth]{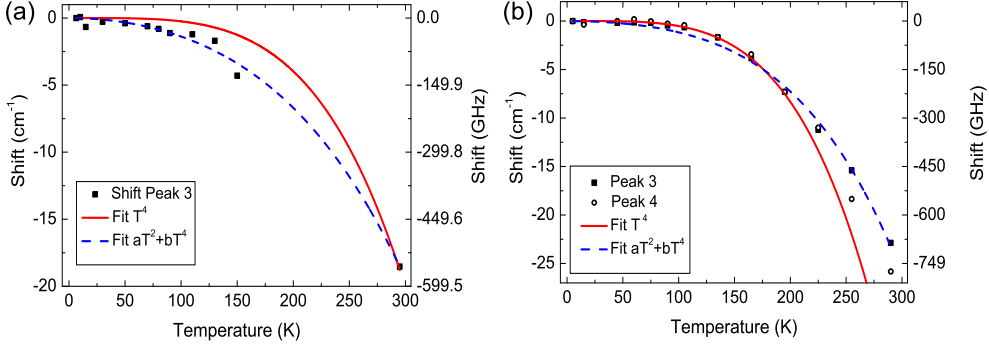}
\caption{ZPL line shift (a) for the ensemble of SiV centres (b) for single SiV centre (emitter C2). The peaks are consecutively numbered with increasing wavelength. Note that the fit takes into account the data of peak 3. Note also that the data points between 150 K and 300 K in (a) are missing due to technical problems during the measurements.  \label{Fig_lineshift}}
\end{figure}
For the application of SiV centres as sources of indistinguishable photons as well as for the coupling of SiV centres to resonant photonic structures, knowledge of the precise temperature dependent wavelength of the ZPL is crucial. All observed SiV ZPL fine structure components show a similar blue shift upon cooling. Figure \ref{Fig_lineshift} displays the shift for one peak of emitter C2 as well as for the ensemble in homoepitaxial CVD diamond. The overall shift ranges from -19 cm$^{-1}$ to -31 cm$^{-1}$. Thus, the observed blue shift is in accordance with reports in the literature (-26 cm$^{-1}$, \cite{Gorokhovsky1995}; -22 cm$^{-1}$, \cite{Feng1993}). Note that we choose intense peaks for the analysis as observation of the weak fine structure components is precluded at elevated temperatures; furthermore, the coalescence of peaks leads to uncertainties at elevated temperatures.

Blue shifting of the ZPL upon cooling has also been observed for other colour centres, e.g., NV centres \cite{Hizhnyakov2002,Davies1974}. The cause of the lineshift as well as the exact temperature dependence are discussed controversially in the literature: During cooling the diamond lattice contracts (for recent measurements of the temperature dependent lattice constant see \cite{Stoupin2011}). Lattice contraction as well as temperature dependent electron lattice interaction lead to a temperature dependent bandgap in diamond \cite{Varshni1967}.
Feng et al.\ \cite{Feng1993} relate the ZPL shift to a change in the bandgap $E_G$ of diamond (for details see \cite{Feng1993}). However, the change of $E_G$ measured for diamond \cite{Varshni1967} is not directly proportional to the ZPL shift observed here. Thus, we conclude that the change of $E_G$ cannot be straightforwardly used to describe the temperature dependence of the ZPL shift without making further assumptions.

Lattice contraction furthermore introduces lineshifts similar to shifts due to mechanical stress.  The ZPL shift $\Delta E_l$ induced by lattice contraction can be calculated by \cite{Davies1981}
\begin{equation}
\Delta E_l=A(c_{11}+2c_{12}) \int_0^T \alpha_v(T) dT \label{eq_latticelineshift}
\end{equation}
where $c_{11}$ and $c_{12}$ are the elastic moduli of diamond, $\alpha_v$ is the volume expansion coefficient of diamond and $A$ is the shift rate of the ZPL under hydrostatic compression. As the parameter $A$ is not reported in the literature for the SiV centre, we cannot quantify $\Delta E_l$. Reference \cite{Stoupin2011} finds empirically that below approximately\ 100 K the temperature dependence of $\alpha_v$ matches a $T^3$ dependence, leading to a $T^4$ dependence of $\Delta E_l$. Figure \ref{Fig_lineshift} shows the $T^4$ dependence fitted to the experimental data. It reasonably well describes the temperature dependent shift for the single emitter but shows deviations at temperatures above 200 K whereas the shift for the ensemble is only moderately described. Note the larger error of the shift at elevated temperature due to the coalescence of line components.

As discussed in section \ref{homogeneous_width}, the investigated SiV centres exhibit a temperature dependent homogenous linewidth. Thus, a quadratic electron-phonon coupling is present \cite{Davies1981} that also contributes to the lineshift.
The overall lineshift is given as the sum of the phonon dependent parts and the lattice contribution $\Delta E_l$. Without knowledge of the parameter $A$, we cannot separate the lineshift due to phonon coupling and lattice contraction. The ZPL shift due to quadratic electron-phonon coupling  has been derived to be proportional also to $T^4$ for temperatures well below the Debye temperature $T_D$ of the solid \cite{Maradudin1966} ($T_D \approx 1900$ K for diamond \cite{Hizhnyakov2002}).  Additionally, in analogy to the temperature dependence of the linewidth, one can use the one-phonon absorption spectrum  together with the phonon population to calculate the line shift. Due to  the same line of argument as for the line broadening, using this approach is not feasible to evaluate our data.
Furthermore, Hizhnyakov and coworkers \cite{Hizhnyakov2002} predict a $aT^2+bT^4$ ($a,b=const.$) dependence when taking into account the softening of bonds in the excited state (see also discussion in section \ref{homogeneous_width}). We also fit a $aT^2+bT^4$ dependence to the measured data which more closely fits the results as compared to the $T^4$ model (see figure \ref{Fig_lineshift}). Nevertheless, as Hizhnyakov and coworkers do not comment on the influence of the lattice expansion in their work, this model does not allow for the separation of lattice and phonon contribution either.
Summarizing, the measured ZPL shift is consistent for different emitters and fine structure components and the overall shift agrees with previous observations on SiV centres. The measured data is consistent with a $aT^2+bT^4$ temperature dependence. However, we cannot identify the individual contribution of lattice contraction and phonon coupling to the shift.
\subsection{Temperature dependent population dynamics \label{population_dyn}}
We here deduce the full population dynamics of single SiV centres at different temperatures from excitation power dependent measurements of the $g^{(2)}$ function as discussed below. To estimate the temperature dependence of the dynamics, we perform measurements at room temperature and cryogenic temperature (20--30 K). The $g^{(2)}$ measurements prove single emitter nature for all observed SiV centres. The brightness of single emitters is deduced from saturation measurements.
The observed maximum photon rates $I_{\infty}$ and saturation powers $P_{sat}$ are summarized in table \ref{tabsattempdep} (for a general discussion of saturation curves see, e.g., \cite{Neu2012a}). Note that the values at low temperature have been corrected for a lower excitation and collection efficiency of the confocal setup due to imaging through the cryostat window. $I_{\infty}$ for emitters C1 and C3 agrees within less than 10\% at room and cryogenic temperature. $I_{\infty}$ for emitter C4 changes considerably, indicating a higher count rate at low temperature. However, we point out that for emitter C4, background fluorescence had to be taken into account and was especially pronounced at room temperature. Thus, over-accounting for the background might explain an underestimated value of $I_{\infty}$ at room temperature.

In the literature, temperature dependent non-radiative quenching of SiV centres has been reported: Feng et al.\ \cite{Feng1993} report a decrease of SiV luminescence in polycrystalline diamond samples by a factor of 8 between 50 K and room temperature and infer an activation energy for luminescence quenching of approximately 70 meV. Sternschulte et al.\ \cite{Sternschulte1994} find that the excited state lifetime of SiV centres in bulk homoepitaxial diamond is almost constant for temperatures between 5 K and 150 K and decreases from 4 ns to 2.7 ns upon heating to 300 K. For polycrystalline material, the same work specifies a lifetime shortening from 1.4 ns to 0.8 ns. The findings summarized above indicate a strong influence of the diamond host material on temperature dependent processes. Both reports speculate that non-radiative decay channels due to impurities and crystal defects acting as non-radiative recombination centres or due to stress quenching are responsible for the observed luminescence quenching. Another possible effect modifying luminescence rates at low temperatures is the deactivation of thermally induced transitions emptying metastable shelving states which e.g.\ has been observed for the NV centre \cite{Drabenstedt1999a}.
It should be noted that the mechanism responsible for the temperature dependent quenching has not been identified yet.  For single emitters, as investigated here, the maximum obtainable photon rates at low temperatures might depend both on freezing of thermally activated de-shelving transitions as well as on the opening or suppression of non-radiative decay channels of the individual emitters depending on the local environment. 
 Room-temperature investigations of the radiative quantum efficiency of single SiV centres in the same samples (2a and b) indeed indicate a strong influence of non-radiative decay and a strong variation in quantum efficiencies among individual emitters \cite{Neu2012a}. From our low-temperature measurements discussed here, however, we infer that no significant de-activation of non-radiative channels occurs down to 20 K. This conclusion is also supported by the temperature dependent $g^{(2)}$ measurements discussed below. In the light of the non-identified mechanism of temperature dependent quenching as well as its strong material dependence, we are unable to determine the cause of the temperature independent fluorescence of the single emitters investigated here. The observed saturation powers $P_{sat}$ are of the same order of magnitude for room and cryogenic temperature. Thus, we exclude any significant influence of the temperature on the off-resonant excitation process.
\begin{table}[h]
\centering
\begin{tabular}{|c|c|c|c|c|c|c|c|}
\hline
Emitter & $P_{sat}$ RT & $I_{\infty}$ RT & T$_{low}$& $P_{sat}$  & $I_{\infty}$ &$\Delta P_{sat}$ &$\Delta I_{\infty}$  \\
\hline
C1 &73 $\mu$W &2.39 Mcps& 25 K &97 $\mu$W &2.17 Mcps&-33\%&9\%\\
C3 &65 $\mu$W &0.775 Mcps& 20 K &57 $\mu$W &0.713 Mcps&13\%&7\%\\
C4 &32 $\mu$W &0.593 Mcps& 30 K &18 $\mu$W &1.08 Mcps&45\%&-82\%\\
\hline
\end{tabular}
\hspace{1cm}
\caption{Saturation power $P_{sat}$ and maximum photon count rate $I_{\infty}$ for emitters C1, C3, C4 at room temperature and at cryogenic temperature. The temperature for the cryogenic measurements is given in column T$_{low}$.
$\Delta P_{sat}$ and $\Delta I_{\infty}$ give the deviation of the low temperature value from the room temperature value in \%, calculated from $\Delta I_{\infty}={[I_{\infty}(RT)-I_{\infty}(T_{low})]}/{I_{\infty}(RT)}$ and analogous for $\Delta P_{sat}$.\label{tabsattempdep}}
\end{table}
\begin{figure}[h]
\centering
\includegraphics[width=\textwidth]{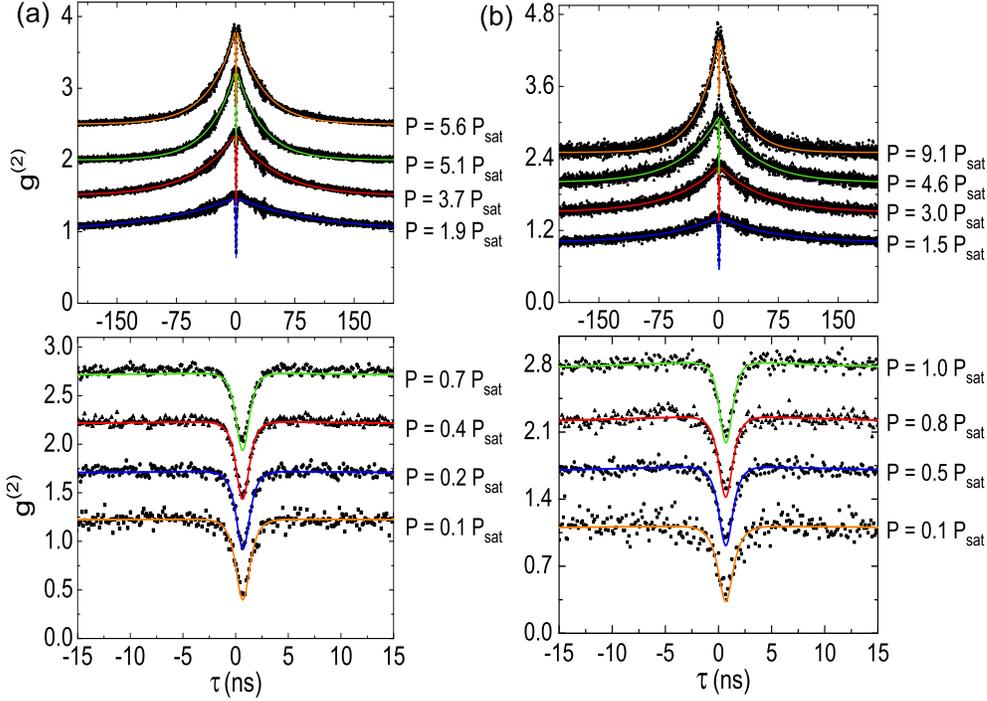}
\caption{Excitation power dependent $g^{(2)}$ functions for emitter C1 (a) at room temperature and (b) at 30 K. he upper part of each graph shows a large range of delay times $\tau$ to enable observation of the bunching; the lower part displays a reduced range of $\tau$ to reveal the antibunching in detail.   \label{Fig_g2functions}}
\end{figure}

Figure \ref{Fig_g2functions} exemplarily displays the excitation power dependent $g^{(2)}$ functions  for emitter C1 at room temperature and 30 K.
The measured $g^{(2)}$ data are fitted with the $g^{(2)}$ function of a three level system
\begin{equation}
g^{(2)}(\tau)=1-(1+a)\,e^{-|\tau|/\tau_1}+a\,e^{-|\tau|/\tau_2}
\label{g23level}
\end{equation}
which we furthermore modify to take into account the instrument response and possible background fluorescence (for details see \cite{Neu2011,Neu2012a}). The employed three level system, described in detail in \cite{Neu2011,Neu2012a}, takes into account two levels for the ZPL transition (1 and 2) and a shelving state (3). The $g^{(2)}$ function reflects the population dynamics of state 2 and can be obtained via solving the rate equations for the populations in the system \cite{Neu2011,Neu2012a}. It includes three parameters $a$, $\tau_1$ and $\tau_2$ that determine the characteristics of the $g^{(2)}$ function: $\tau_1$ governs the antibunching dip, whereas $\tau_2$ gives the time constant of the bunching and $a$ determines how pronounced the bunching is. The fits agree well with the measured data for all delay times $\tau$ and identify all SiV centres as single emitters. The deviation of $g^{(2)}(0)$ from 0 is induced by the instrument response of the measurement setup as confirmed by the accordance of the fitted $g^{(2)}$ functions and the measured data. The deviation $\Delta g^{(2)}(0)$ between the fitted value of $g^{(2)}(0)$ and the measured datapoints is below 0.07 for emitter C1 and below 0.15 for emitter C3 without any background correction and is mainly governed by the noise of the $g^{(2)}$ function.  For emitter C4, $\Delta g^{(2)}(0)=0.04$ is obtained after background subtraction.

Treating the population dynamics of single SiV centres in the framework of a three level system marks a significant simplification as we observe four radiative transitions at low temperature. Our $g^{(2)}$ measurements correlate photons from all four different transitions due to broadband (20 nm) spectral filtering. Thus, we measure an 'effective' $g^{(2)}$ function with properties dominated by the fine structure lines with high relative intensity. The full conformity of the measured $g^{(2)}$ data with the fitted three level $g^{(2)}$ function justifies the description as a three level system and excludes, e.g., that several shelving states with different lifetimes have to be considered which would induce additional time constants \cite{Boiron1996}. However, we note that detailed consideration of the line intensities in the temperature dependent fluorescence spectra (method see \cite{Clark1995}) furthermore indicates that population transfer takes place in the excited state (thermalization). Such processes are also not taken into account by the simplified three level model.

\begin{figure}[h]
\centering
\includegraphics[width=\textwidth]{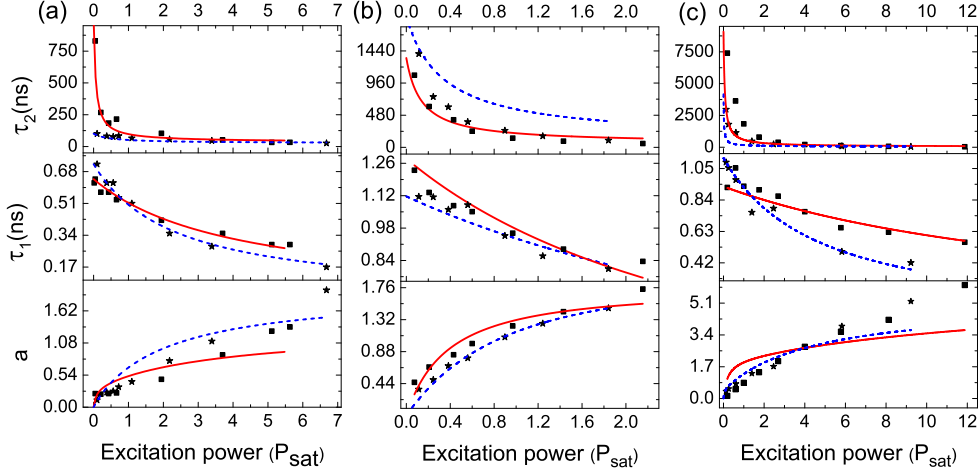}
\caption{$g^{(2)}$ parameters $a$, $\tau_1$ and $\tau_2$ for (a) emitter C1, (b) emitter C3, (c) emitter C4,  at room temperature and cryogenic temperature (30K C1,C4; 20 K, C3). Filled squares: data points at room temperature (Fit red curve), filled stars: cryogenic temperature (Fit blue, dashed curve). For discussion of the employed population dynamics model see text. \label{Fig_g2par}}
\end{figure}
From the fits of the $g^{(2)}$ functions, we obtain the power dependent parameters $a$, $\tau_1$ and $\tau_2$ in (\ref{g23level}) summarized in figure \ref{Fig_g2par}. The parameter range of $a$, $\tau_1$ and $\tau_2$ is in accordance with previous room temperature observations \cite{Neu2011,Neu2012a}; the comparison of room temperature and low temperature data indicates mostly temperature independent population dynamics. The rate coefficients $k_{ij}$ (transition from state $i$ to $j$) fully describe the internal population dynamics of the SiV centres. In previous work \cite{Neu2011,Neu2012a}, we developed a model for the internal dynamics: $k_{21}$ and $k_{23}$ are intensity independent, $k_{12}$ depends linearly on the excitation power $P$ ($k_{12}=\sigma P$) and the de-shelving rate coefficient $k_{31}$ is given as
\begin{equation}
k_{31}=\frac{d\cdot P}{P+c}+k_{31}^0. \label{k31}
\end{equation}
Equation (\ref{k31}) introduces the possibility that the colour centre is transferred from the shelving state (state 3) to the ground state (state 1) as a consequence of laser excitation (intensity dependent de-shelving), in addition to the spontaneous relaxation to state 1 represented by $k_{31}^0$. Figure \ref{Fig_g2par} summarizes the power dependent curves for $a$, $\tau_1$ and $\tau_2$ obtained in this model. As discernible from figure \ref{Fig_g2par}, the model suitably describes the power dependence of $a$, $\tau_1$ and $\tau_2$ with the exception of $\tau_2$ for emitter C3. Thus, an intensity dependent de-shelving process is present at room and cryogenic temperature.

To estimate the rate coefficients in this model, we obtain the limiting values $\tau_1^0$, $\tau_2^0$, $\tau_2^{\infty}$ and $a^{\infty}$ for vanishing and infinite excitation power from the data given in figure \ref{Fig_g2par}. For details on the model and the procedure to determine the rate coefficients see \cite{Neu2011,Neu2012a}.  The resulting rate coefficients are summarized in table \ref{cryoRTrates}.
\begin{table}
\centering
\begin{tabular}{|c|c|c|c|c|c|c|}
\hline
\textbf{Em C1} & $k_{21}$(MHz) & $k_{23}$(MHz)& $k_{31}^0$(MHz) &$d$ (MHz) &$\sigma$(MHz/$\mu$W)& $c$ ($\mu$W) \\
\hline
RT & 1545.1  & 17.4  & 1.0 & 11.9  & 5.21  & 80.4  \\
\hline
30 K & 1363.9  & 25  & 10  & 2.79  & 6.3  & 60  \\
\hline
\hline
\textbf{Em C3} & $k_{21}$(MHz) & $k_{23}$(MHz)& $k_{31}^0$(MHz) &$d$ (MHz) &$\sigma$(MHz/$\mu$W)& $c$ ($\mu$W) \\
\hline
RT & 770.1  & 11.1  & 0.749 & 5.65  & 3.79  & 217  \\
\hline
20 K & 889.5  & 5.73  & 0.513  & 3.3  & 3.1  & 558  \\
\hline \hline
\textbf{Em C4} & $k_{21}$(MHz) & $k_{23}$(MHz)& $k_{31}^0$(MHz) &$d$ (MHz) &$\sigma$(MHz/$\mu$W)& $c$ ($\mu$W) \\
\hline
RT & 1053.6  & 21.7  & 0.11 & 3.44 & 1.83  & 201  \\
\hline
30 K & 874.8  & 18.8 & 0.24 & 3.4 & 11.0 & 8.1  \\
\hline
\end{tabular}
\caption{Rate coefficients for emitter C1, C3 and C4 deduced from the limiting values of $a$, $\tau_1$ and $\tau_2$ using the three level model including intensity dependent de-shelving.  \label{cryoRTrates}}
\end{table}
$k_{21}$ is mainly determined by the limiting value $\tau_1^0$ which is an estimate for the excited state lifetime of the colour centre. As the deviation of low temperature and room temperature values of $k_{21}$ is smaller than 20 \% without a clear trend, we conclude that no temperature dependent activation or de-activation of non-radiative channels occurs in accordance with the observed brightness of the emitters.

The emitters reveal significantly different shelving dynamics represented by the rate coefficients $k_{23}$, $k_{31}^0$ and $d$. However, the room and low temperature parameters for the same emitter are in the same order of magnitude, thus the shelving dynamics do not fundamentally change upon cooling. For emitter C1, $k_{31}^0$ is inconsistent at high and low temperature. However, we note that this might be connected to the challenging determination of $\tau_2^0$ which  defines $k_{31}^0$: at low temperature, the lowest excitation power used was $0.1 P_{sat}$; at room temperature  $0.02 P_{sat}$ could be employed due to the increased performance of the confocal setup at room temperature, and thus the reduced measurement time. The observation of a temperature independent shelving dynamics excludes thermally activated shelving or de-shelving processes. This finding is in contrast to observations on NV centres for which the shelving state is found only 37 meV below the excited state and the centres can be thermally re-excited from the metastable shelving state at elevated temperature \cite{Drabenstedt1999a}.
\section{Conclusion}
We use cryogenic temperature luminescence spectroscopy to investigate single SiV centres in NDs on Ir as well as an ensemble of SiV centres in a high quality, low stress homoepitaxial CVD diamond film. As a result of stress in the NDs, the observed individual fine structure splittings for single emitters vary significantly. However, we find no discernible trend of the modifications. We measure the temperature dependent homogeneous broadening which is best described by a $T^3$ dependence and might be associated to phonon-broadening in the presence of impurities. Spectral diffusion additionally broadens the emission lines to 25--160 GHz at around 5 K for most of the single emitters. In contrast, the ensemble in the homoepitaxial CVD film shows a linewidth of individual fine structure components of down to 9 GHz.  The line position of the ZPL is found to blue shift by 20-30 cm$^{-1}$ upon cooling from room temperature to 5 K. The temperature dependence of the shift is consistent with a $aT^2+bT^4$ dependence and is related to quadratic electron-phonon coupling and lattice contraction. However, due to the undetermined stress response of SiV centres, we cannot infer the dominant mechanism. All emitters reveal population dynamics that are reasonably described in the framework of a three level intensity dependent de-shelving model at cryogenic as well as room temperature. The population dynamics is mostly temperature independent without de-activation of non-radiative processes or changes in shelving or de-shelving dynamics.

A number of recent investigations have shown that SiV centres contained in NDs are very promising single photon emitters \cite{Neu2011,Neu2012a}. In view of the generation of indistinguishable single photons, emitters with narrow linewidths and narrow inhomogeneous distribution of ZPL positions are required. A possible route to combine the superior spectral emission properties of the homoepitaxial diamond film as demonstrated here with the high collection efficiency from nanodiamonds \cite{Neu2012a} is the fabrication of photonic structures, e.g. diamond nanowires \cite{Babinec2010}.
\section*{Acknowledgements}
The project was financially supported by the Bundesministerium f\"ur Bildung und Forschung within the projects EphQuaM (contract 01BL0903) and QuOReP (contract 01BQ1011). The SEM-analysis of the samples was performed by J. Schmauch  (Universit\"at des Saarlandes, Saarbr\"u\-cken, Germany).

\end{document}